\documentclass[useAMS, usenatbib]{mn2e}
\usepackage{times}
\usepackage{natbib}
\usepackage{graphicx}
\usepackage{amsmath}
\usepackage[T1]{fontenc}
\usepackage{aecompl}

\newcommand{\str}{Str\"{o}mgren}

\newcommand{\nH}{n_{\rm H, \infty}}
\newcommand{\nseven}{10^7~{\rm cm^{-3}}}

\newcommand{\none}{10~{\rm cm^{-3}}}
\newcommand{\lambdaavg}{\langle \lambda_{\rm rad} \rangle}

\newcommand{\hi}{\rm H~{\textsc i}}
\newcommand{\hii}{\rm H~{\textsc {ii}}}
\newcommand{\hei}{\rm He~{\textsc {i}}}
\newcommand{\heii}{\rm He~{\textsc {ii}}}
\newcommand{\heiii}{\rm He~{\textsc {iii}}}
\newcommand{\lmone}{l \propto \dot{m}}
\newcommand{\lmtwo}{l \propto \dot{m}^2}

\newcommand{\msun}{{\rm M}_\odot}

\title[Compton heating in accretion on to BHs]
{The role of Compton heating in radiation-regulated accretion on to black holes}
\author[K. Park, M. Ricotti, T. Di Matteo, and Reynolds]{KwangHo
Park$^{1}$\thanks{E-mail: kwanghop@andrew.cmu.edu}, 
Massimo Ricotti$^{2}$, Tiziana Di Matteo$^{1}$, and Christopher S. Reynolds$^{2}$ 
\\ 
$^{1}$McWilliams Center, Carnegie Mellon University, Pittsburgh,
PA 15213, USA\\
$^{2}$Department of Astronomy, University of Maryland, College Park,
MD 20740, USA}

\begin{document}
\date{Accepted. Received 2014}

\pagerange{\pageref{firstpage}--\pageref{lastpage}} \pubyear{2014}

\maketitle

\label{firstpage}

\begin{abstract}
We investigate the role of Compton heating in radiation-regulated
accretion on to black holes from a neutral dense medium using 1D
radiation-hydrodynamic simulations. We focus on the relative effects
of Compton-heating and photo-heating as a function of the spectral
slope $\alpha$, assuming a power-law spectrum in the energy range
of $13.6$~eV--$100$~keV. While Compton heating is dominant only
close to the black hole, it can reduce the accretion rate to $0.1$\%
($\lmtwo$ model)--$0.01$\% ($\lmone$ model) of the Bondi accretion
rate when the BH radiation is hard ($\alpha \sim 1$), where $l$ and
$\dot{m}$ are the luminosity and accretion rate normalised by Eddington
rates, respectively.  The oscillatory behaviour otherwise typically
seen in simulations with $\alpha > 1$, become suppressed when $\alpha
\sim 1$ only for the $\lmone$ model. The relative importance of the
Compton heating over photo-heating decreases and the oscillatory
behaviour becomes stronger as the spectrum softens. When the spectrum
is soft ($\alpha > 1.5$), photo-heating prevails regardless of
models making the effect of Compton heating negligible. On the
scale of the ionization front, where the gas supply into the
\str~sphere from large scale is regulated, photo-heating dominates.
Our simulations show consistent results with the advection-dominated
accretion flow ($\lmtwo$) where the accretion is inefficient and
the spectrum is hard ($\alpha \sim 1$).  
\end{abstract}



\section{Introduction}
\label{sec:intro}
A realistic estimation of an accretion rate for a black hole (BH)
is of great cosmological importance.  It does not only determine
its cosmological growth rate of the BH, but also it is linked directly to the
BH luminosity, and thus radiative-feedback effects on the neighbouring gas of the
host galaxies and intergalactic medium. Commonly, the
Eddington-limited Bondi--Hoyle rate \citep{BondiH:44,Bondi:52} is
used as the prescription for the accretion rates of the BHs in
cosmological simulations.  However, this simple BH accretion recipe
is not physically motivated since the gas accretion on to BHs is
regulated by radiative and mechanical feedback from the BH. Due to
the feedback, the thermal and dynamical structure of the accretion
flow is far more complicated than the one described by the classical
Bondi accretion calculation. The radiative feedback by BHs is important when
the accretion activity of the BH peaks (i.e. quasar mode of active
galactic nuclei, AGN), whereas the mechanical feedback is prominent
in the low-accretion state (i.e., radio mode of AGN). Additionally,
the BH spectrum in high-energy X-ray is known to display different
characteristics switching between {\it hard} and {\it soft} states
depending on the accretion luminosity
\citep{Miyamoto:92,Miyamoto:93,Belloni:11,BegelmanA:14}. Thus, 
understanding how the transitions of the BH spectrum are coupled
with the accretion luminosity is important not only for understanding
the physics of the accretion flow but also for estimating the
accretion rate.

High-energy UV and X-ray photons emitted from BHs regulate the gas
supply to the BH by heating and ionizing the neighbouring gas
\citep{OstrikerWYM:76,CowieOS:78,Shull:79}, as well as imposing
direct radiation forces on the gas inflow
\citep{OstrikerCCNP:10,Choi:2012,Choi:2014}.   This induces a
self-regulated feedback loop between the inflowing 
gas and the outflowing radiation. A series of
papers by \citet{ParkR:11,ParkR:12,ParkR:13} focused on the role
of photo-ionization and photo-heating for BHs with a mass $M_{\rm
bh}$ accreting from the neutral, warm ($T_\infty \sim 10^4$~K), and
dense ($\nH (M_{\rm bh}/100~\msun)$ = $10^5$--$\nseven$) gas where
$M_{\rm bh}$ is the BH mass and $\nH$ is the gas number
density. In this particular scenario, motivated by cosmological
simulations of the early universe, a well-defined hot ionized bubble
of gas forms around the BH that is surrounded by an optically
thick and neutral gas reservoir. The ionization front (I-front) formed
by high-energy photons, which is typically a few orders of magnitude
larger than the classical Bondi radius ($r_{\rm B}=GM_{\rm bh}/c_{s,
\infty}^2$, where $c_{s, \infty}$
is the sound speed of the gas), plays a critical role in regulating
gas supply to the BH when the BH is at rest \citep{ParkR:11,
ParkR:12} or in supersonic motion \citep{ParkR:13}. Pressure
equilibrium between the neutral and ionized gas across the I-front is
the key factor for explaining reductions in the accretion rate
\citep{ParkR:11,ParkR:12,ParkR:13}. Thus, understanding the heating
mechanisms or the instabilities
\citep{Williams:99,Williams:02,Mizuta:2005,WhalenN:08a,WhalenN:08b,WhalenN:11,ParkRDR:2014,
Ricotti:2014} inside the \hii~region and near the I-front that might
modify the thermal structure of the \str~sphere is extremely
important in understanding the radiation-regulated growth of the BHs.

In the physical conditions close to an accreting BH, the strong radiation
field and the (ionized) gas can exchange energy directly through
Compton scattering.  The direction of energy exchange is such as to
bring the radiation-field and gas closer to thermal equilibrium. The 
subsequent Compton heating (or cooling) process increases (decreases) the gas temperature
to the characteristic {\it Compton temperature} $T_C \sim 2 \times
10^7~$K \citep{Sazonov:2004}. Due to the high $T_C$ characterizing 
BH accretion, Compton heating may be an important radiative feedback 
mechanism for the super-massive black holes (SMBHs) at the centres of elliptical galaxies, 
inducing strong
intermittencies of the BH luminosity \citep{Sazonov:05, CiottiO:01,
CiottiO:07, CiottiOP:09, NovakOC:11, NovakOC:12, Choi:2012, Choi:2014}.
However, Compton heating has been commonly neglected for the case
of intermediate mass black holes (IMBHs) in the mass range of
$10^2$--$10^4~\msun$ \citep{MiloCB:09,Li:11,ParkR:11,ParkR:12,ParkR:13}
where photo-heating is assumed to be dominant. Understanding the
role of Compton heating across the BH mass range is important
for building a universal BH growth model that connects
IMBH and SMBHs \citep{Ferrara:2014,Natarajan:2014}. 
Both the Compton and photo-heating rates show the same dependence on
the distance from the BH being proportional to the radiation flux,
while they show different dependence on the ionization fraction of
the medium \citep[e.g.,][]{MadauE:99,CiottiO:01}. The neutral
fraction of the hydrogen inside a \hii~region increases as a function
of radius; however, the fraction of the ionized atoms is close to unity
since most of the hydrogen and helium is ionized. 
This fact suggests that photo-heating can be relatively more efficient
in heating the gas near the I-front.

In this study, we focus on the role of Compton heating relative to
photo-ionization and photo-heating for the accretion
on to BHs embedded in an optically thick neutral medium. The primary
goal of this paper is to explore when one of the two heating
mechanisms dominates over the other as a function of the hardness
of the spectrum. In Section~\ref{sec:method}, we describe the
numerical simulation setup.  We present our results in
Section~\ref{sec:results}. In Section~\ref{sec:summary}, we summarize
and discuss our findings.

\section{Methodology}
\label{sec:method}
\subsection{Compton heating as a function of spectral index}
\label{sec:compton}
The Compton heating (or cooling) rate is determined by the number of
scatterings $N_{\rm scatt} (\nu)$ per unit time, and the energy exchange per
interaction $\Delta E (\nu , T)$, between the high-energy photons
from a BH and the electrons in the neighbouring gas medium with a gas
temperature $T$. Under the assumption of spherical symmetry, 
the Compton heating rate ($\Gamma_C$) is given by: 
\begin{equation}
\Gamma_C = \frac{\sigma_{\rm T} n_{\rm e}}{m_{\rm e} c^2} 
\int_{\nu_{\rm min}}^{\nu_{\rm max}} d\nu F_\nu (h\nu - 4 k_B T),
\label{eq:compt_integral}
\end{equation} 
where $n_{\rm e}$ is the electron number density, and $\sigma_{\rm
T}$ is the Thomson cross-section. The radiation flux $F_\nu$ at 
frequency $\nu$ in Equation~(\ref{eq:compt_integral}) depends on
the distance from the BH and the optical depth $\tau_\nu$, however
the optical depth can be generally neglected in optically-thin situations
($F_\nu = L_\nu /4\mathrm{\pi} r^2$) such as in the vicinity of the BH.
Assuming a power-law spectrum ($F_\nu =C \nu^{-\alpha}$), the
radiation flux can be expressed $F=C (\nu_{\rm max}^{1-\alpha}-
\nu_{\rm min}^{1-\alpha})/(1-\alpha)$ for $\alpha \ne 1$ while $F=C
{\rm ln} (\nu_{\rm max}/\nu_{\rm min})$ for $\alpha=1$.
Equation~(\ref{eq:compt_integral}) can be integrated to give a
simple expression:
\begin{equation}
\Gamma_C = F \frac{\sigma_{\rm T} n_{\rm e}}{m_{\rm e} c^2} \left(  \frac{h\nu_{\rm max}}{\Lambda_C}-4 k_{\rm B} T \right),
\label{eq:compt_simple}
\end{equation}
where $\Lambda_C = {\rm ln} (\nu_{\rm max}/\nu_{\rm min})$ for the
special case of $\alpha=1$. In general, $\Lambda_C$ can be expressed
for $ 1 < \alpha < 2$ as
\begin{equation}
\Lambda_C =  \left(\frac{2-\alpha}{\alpha-1}\right)
\frac{\nu_{\rm max}^{2-\alpha}- \nu_{\rm min}^{2-\alpha} (\nu_{\rm max}/\nu_{\rm min}) } {\nu_{\rm
max}^{2-\alpha}- \nu_{\rm min}^{2-\alpha}}.  
\label{eq:lambda_c}
\end{equation}

\begin{figure}
\includegraphics[width=85mm]{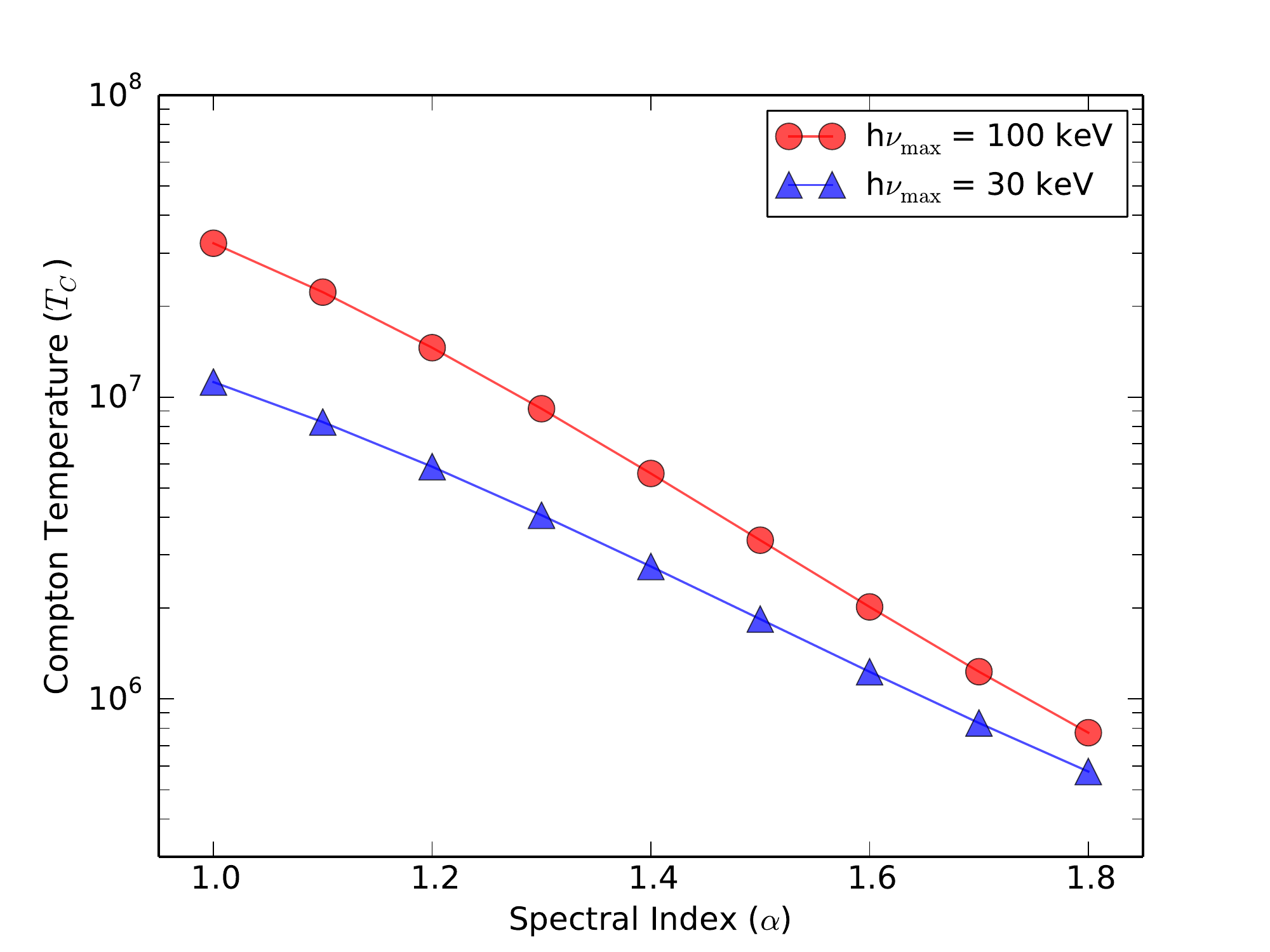}
\caption{The Compton temperature $T_C$ as a function of spectral index
($\alpha$). Circles and triangles show $T_C$ for the maximum energy
of the photon $E_{\rm max}=100$ and $30$~keV, respectively. With
a fixed $E_{\rm max} = 100$~keV, the Compton temperature is $T_C \sim 3\times
10^7~$K for $\alpha = 1$ and monotonically decreases with  
increasing $\alpha$, reaching $T_C \sim 3\times 10^6~$K for $\alpha
\sim 1.5$. With $E_{\rm max} = 30$~keV, the Compton temperature is
$T_C \sim 10^7~$K at $\alpha = 1$, which is lower than the $T_C$
for $E_{\rm max} = 100~$keV. The difference between $T_C$ for
different $E_{\rm max}$ becomes reduced with increasing $\alpha$
due to the extra dependence on $E_{\rm max}$ and $\alpha$ shown in
Equation~(\ref{eq:lambda_c}). }
\label{fig:compt_temp} 
\end{figure}

From Equation~(\ref{eq:compt_simple}), the characteristic Compton
temperature $T_C$ is defined as
\begin{equation}
T_C \equiv \frac{h\nu_{\rm max}}{4 k_{\rm B} \Lambda_C}, 
\end{equation}
which is the temperature when the radiation field and the
gas are in equilibrium state. The Compton temperature monotonically
increases with the maximum photon energy $E_{\rm max} = h\nu_{\rm
max}$ and decreases as a function of the spectral index $\alpha$.
In Fig.~\ref{fig:compt_temp}, circles and triangles show the
Compton temperature as a function of the spectral index $\alpha$ for
$h\nu_{\rm max} = 100$ and $30$~keV respectively with the fixed $E_{\rm
min}=13.6~$eV. For example, with $\alpha=1$
and $E_{\rm max}=100~$keV, the Compton temperature reaches $T_C
\sim 3\times 10^7~$K,
while $T_C \sim 3\times 10^6~$K when $\alpha=1.5$. Since $T_C$ is
approximately proportional to $E_{\rm max}$, the $T_C$ is lower
when $E_{\rm max} =30$~keV. The maximum photon energy of $E_{\rm
max}= 100~$keV is adopted since Compton scattering becomes less
efficient for photons with energies above $E \sim 100$~keV,
where the relativistic effects encapsulated by the Klein--Nishina
correction reduce the scattering cross-section $\sigma_{\rm T}$
\citep{KleinN:29, Blumenthal:74}.  In addition, the Nuclear Spectroscopic Telescope
Array (NuSTAR) has shown that spectra of many
(non-blazar) accreting BH systems appear to cut off at approximately 100\,keV
due to the physics of the accretion disk corona \citep{Harrison:2013}.

\citet{ParkR:11, ParkR:12} found that the thermal structure of the
\hii~region (i.e., the temperatures in and out of the \str~sphere)
created by UV and X-ray photons from BHs determines the accretion
rate. They also found that the typical temperature $T_{\rm in}$
at the effective accretion radius inside the \str~sphere is altered 
by the hardness of the emitted radiation (i.e., the spectral index
$\alpha$ of the power-law spectrum). $T_{\rm in}$ increases with
the harder spectrum, but still remains
less than $\sim 10^5~$K. This results from the fact that the
mean energy of the emitted photons increases with decreasing $\alpha$.
Since both of the Compton heating and photo-heating depend strongly
on the hardness of the spectrum, the relative importance of the two
heating mechanisms must be compared with the same spectral index
$\alpha$.

\subsection{Accretion feedback models}
The bolometric luminosity of a BH is $l = \eta \dot{m}$, where the
luminosity and the accretion rate are normalised by the Eddington
rates as $l \equiv L/L_{\rm Edd}$ and $\dot{m} \equiv \dot{M}/\dot{M}_{\rm
Edd}$. The Eddington luminosity is $L_{\rm Edd} = 4\pi GM_{\rm
bh}m_{\rm p}c \sigma_{\rm T}^{-1}$ and we define the Eddington
accretion rate $M_{\rm Edd} \equiv L_{\rm Edd} c^{-2}$. For a thin
disc \citep{ShakuraS:73} model, the radiative-efficiency is constant
$\eta \sim 0.1$ and $0 < \dot{m} < \eta^{-1} \sim 10$ for Eddington-limited
accretion. On the other hand, the efficiency
increases as the accretion rate increases for spherical accretion
or an advection-dominated accretion flow (ADAF), going as $\eta
\propto \dot{m}$ \citep{Shapiro:73,NarayanY:95,Abramowicz:96,ParkO:01}.
The ADAF model
($\lmtwo$) is radiatively-inefficient and is applicable when the accretion rate is low. 
The soft synchrotron photons from the ADAF inverse Compton scatter off the hot ($\sim 100$\,keV)
electrons in the plasma and produce a hard spectrum
($\alpha = 1$) \citep{Narayan:98}. 
Interestingly, it has been shown that the photo-heating/ionization
alone does not produce a significant difference between $\lmone$
and $\lmtwo$ models when a same spectral index $\alpha$ is assumed
for these two models \citep{ParkR:11}. This is due to the fact that
the thermal structure inside the {\hii}~region (described by the parameter
$T_{\rm in}$), the key for determining the accretion rate, is not
affected by the difference between the two models but significantly 
altered by the spectral index as explained in the previous
section. 
In this study, we examine how the inclusion of Compton heating affects
the different accretion models ($l = 0.1\dot{m}$ and $l = 0.1\dot{m}^2$)
focusing on the accretion rate and the oscillatory behaviour found
in the previous works.

\subsection{Radiation-hydrodynamic simulation setup}
We run a suite of 1D radiation-hydrodynamic simulations for a fixed
BH mass $M_{\rm bh} = 10^6$~M$_\odot$ and a gas density $\nH=\none$.
We use the non-relativistic hydrodynamic simulation code {\sc
zeus-mp} \citep{StoneN:92, Hayes:06} with our 1D radiative transfer
subroutine \citep{RicottiGS:01,WhalenN:06}. 
The BH is assumed to be located
at the origin of the computational domain which we cover with $256$--$512$ 
logarithmically spaced radial grid zones.  We
assume spherical symmetry. We adopt an operator-split method that
alternates hydrodynamic and radiative-transfer calculations, using
the smaller of the hydrodynamical and {\rm ionization} time steps at
each time-step $\Delta t$=min$(\Delta t_{\rm hydro}, \Delta t_{\rm
ion}$). The mass accretion rate is not prescribed but instead directly
read at the minimum radius $r_{\rm min}$ at every time-step and
converted to the radiation luminosity depending on the accretion model.  
For the radiative-transfer calculation, the number of photons in 50 logarithmically spaced energy bins from
$13.6$~eV to $100$~keV are calculated using the instantaneous BH luminosity and
the spectral index $\alpha$. Our subroutine solves the 1D radiative
transfer equation by calculating photo-heating/ionization, gas
cooling, radiation pressures, and Compton heating as explained in
Section~\ref{sec:compton}. The Compton heating rate $\Gamma_C$ is
calculated directly by performing the integral shown in
Equation~(\ref{eq:compt_integral}). Our 1D radiative transfer
subroutine updates the abundances of \hi, \hii, \hei, \heii, \heiii,
and $e^-$ at each radius for every time-step. We run simulations
with a different spectral index in the range of $ 1.0 \le \alpha
\le 1.8 $ with a fixed energy range from $E_{\rm min}= 13.6$~eV to
$E_{\rm max}= 100$~keV. The spectral index $\alpha$ is kept 
constant in each simulation.

\section{Results}
\label{sec:results}
\subsection{Accretion rate as a function of spectral index}
\begin{figure}
\includegraphics[width=85mm]{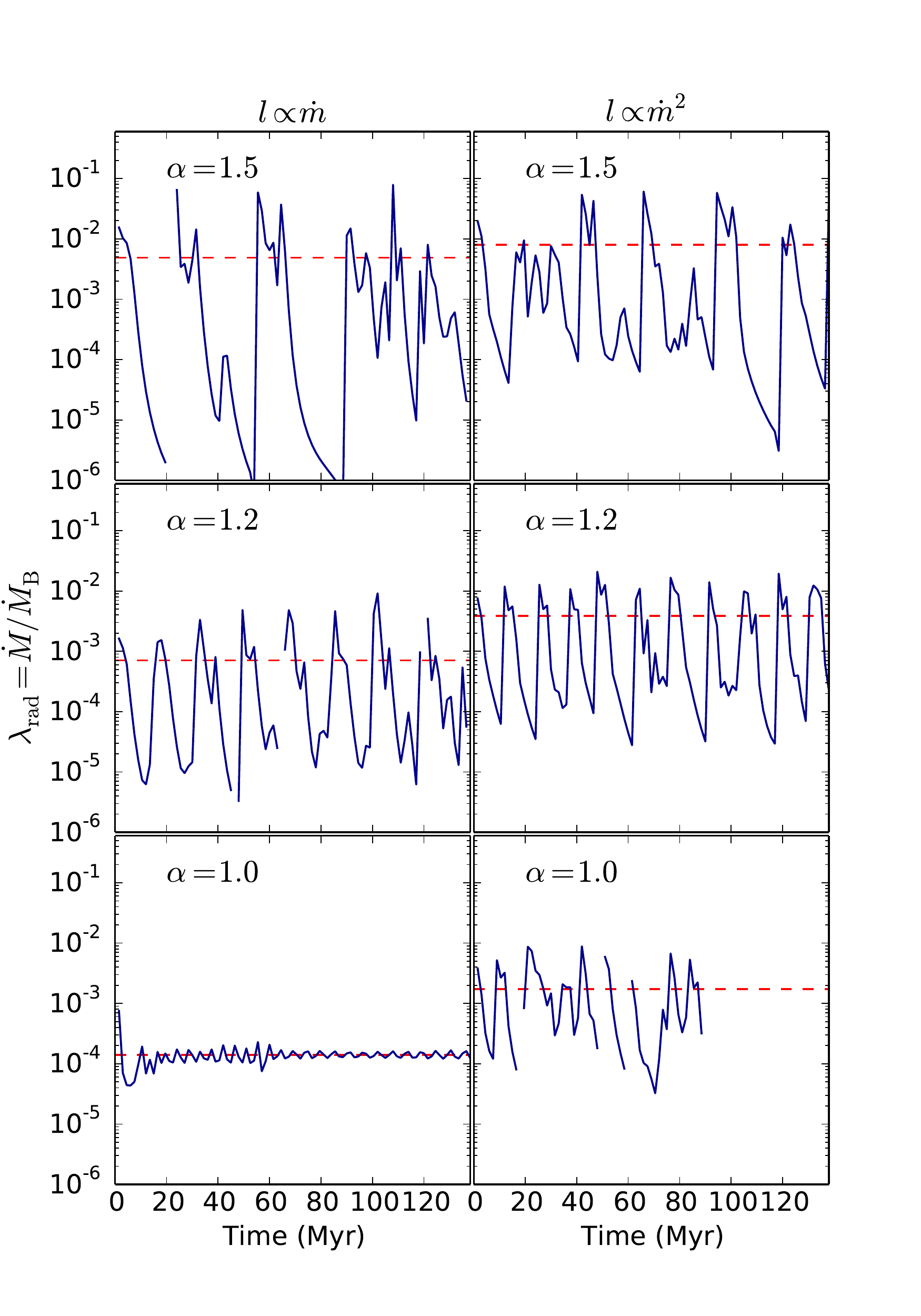}
\caption{Evolution of the accretion rates for simulations with
different spectral index ($\alpha=1.5, 1.2$ and $1.0$) for accretion
models $\lmone$ (left panels) and $\lmtwo$ (right panels).
$\lambda_{\rm rad}$ shown here is the accretion rate normalised by
the Bondi rate ($\dot{M}_{\rm B}$). As the Compton temperature
increases with decreasing spectral index $\alpha$ from top to bottom
panels, the accretion rate and the amplitude of the oscillations
for both models decrease. For the case of $\alpha =1.0$, oscillatory
behaviour is almost suppressed for the $\lmone$ model, while $\lmtwo$
model still displays oscillation of the accretion rate. The $\lmone$
model shows significantly reduced time-averaged accretion rate
$\lambdaavg$, shown as a dashed line in each panel, while the
$\lmtwo$ model displays the milder change.}
\label{fig:mdot_t}
\end{figure}

Fig.~\ref{fig:mdot_t} shows the accretion rates ($\lambda_{\rm rad}
= \dot{M}/\dot{M}_B$) as a function of time for the simulations
with Compton heating included. We normalise the accretion rate
to the Bondi accretion rate $\dot{M}_{B} = \pi e^{3/2} \lambda_{\rm
B} G^2 M_{\rm bh}^2 \rho_\infty c_{s,\infty}^{-3}$
where $\lambda_{\rm B}$ is the dimensionless accretion rate 
assuming isothermal equation of state ($\gamma=1$), $\rho_\infty$ is the
density of the gas. For the assumed simulation parameters ($M_{\rm
bh}\nH = 10^7~\msun {\rm cm}^{-3}$, $T_\infty=10^4$~K, and $\eta=0.1$), it
can be shown that $\dot{m} \sim 2.5 \lambda_{\rm rad}$. The left panels
show the model $\lmone$ while the right panels show the model
$\lmtwo$.  From top to bottom, the spectral index decreases ($\alpha
= 1.5, 1.2,$ and $1.0$). Strong oscillation of the accretion rate
is seen for $\alpha = 1.5$ for both models, however the oscillatory
behaviour becomes suppressed with decreasing spectral index. The
accretion
amplitude between the maximum and the minimum decreases and the
mean accretion rate (shown as dashed lines) also decreases as the
spectrum becomes harder for both $\lmone$ and $\lmtwo$ models. For
$\lmone$ model at $\alpha=1.0$, the oscillation becomes
almost completely suppressed and the mean accretion
rate becomes $\sim 10^{-4}$ of the Bondi rate. Mild oscillations are
observed for the model $\lmtwo$ at $\alpha =1.0$ and the accretion 
rate becomes $\sim 10^{-3}$ of the Bondi rate, higher than
the rate for the $\lmone$ model.

\begin{figure}
\includegraphics[width=85mm]{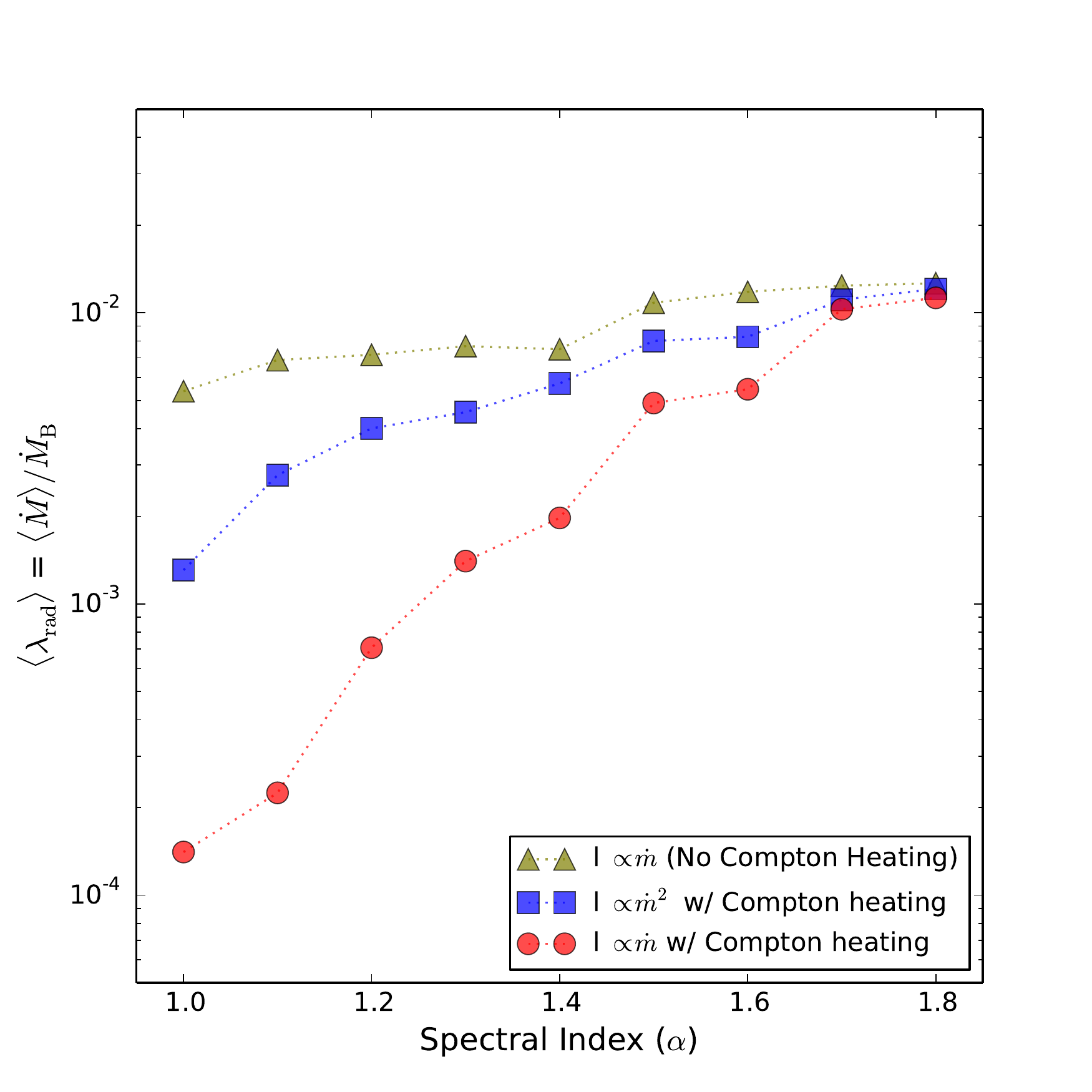} 
\caption{Time-averaged accretion rates normalised by the Bondi rate
$\lambdaavg$ as a function of spectral index in the range $1.0 \le
\alpha \le 1.8$ for different feedback models: $\lmone$ with
photo-heating only (triangles), $\lmtwo$ with Compton heating
(squares), and $\lmone$ with Compton heating (circles). The average
accretion rate for the photo-heating only model shows a mild dependence
on the spectral index while the models with Compton heating show
strong dependence on the spectral index. $\lambdaavg$ is $\sim$
$10^{-4}$ for the model $\lmtwo$ and $\sim 10^{-3}$ for
the model $\lmone$ when $\alpha =1$. The accretion rate for the
model $\lmtwo$ shows intermediate dependence on the spectral index
between the models $\lmone$ and photo-heating only.}
\label{fig:lambda_alpha}
\end{figure}

Fig.~\ref{fig:lambda_alpha} shows the time-averaged accretion rate
$\lambdaavg$ as a function of the spectral index. Different symbols
show different sets of simulations: $\lmone$ with photo-heating
only (triangles), $\lmone$ model with Compton heating (circles),
and $\lmtwo$ with Compton heating (squares).
The average accretion rate for the photo-heating-only model shows
mild dependence on the spectral index, while the both models with Compton
heating show strong dependencies on the spectral
index. $\lmtwo$ model shows that the accretion is suppressed   
as the spectrum becomes harder ( 0.1 per cent of the Bondi rate when $\alpha \sim 1$). 
The average accretion rate for the model $\lmtwo$ shows
milder dependence on the spectral index than the model $\lmone$ does. 

\subsection{Compton heating vs. photo-heating}
\begin{figure}
\includegraphics[width=85mm]{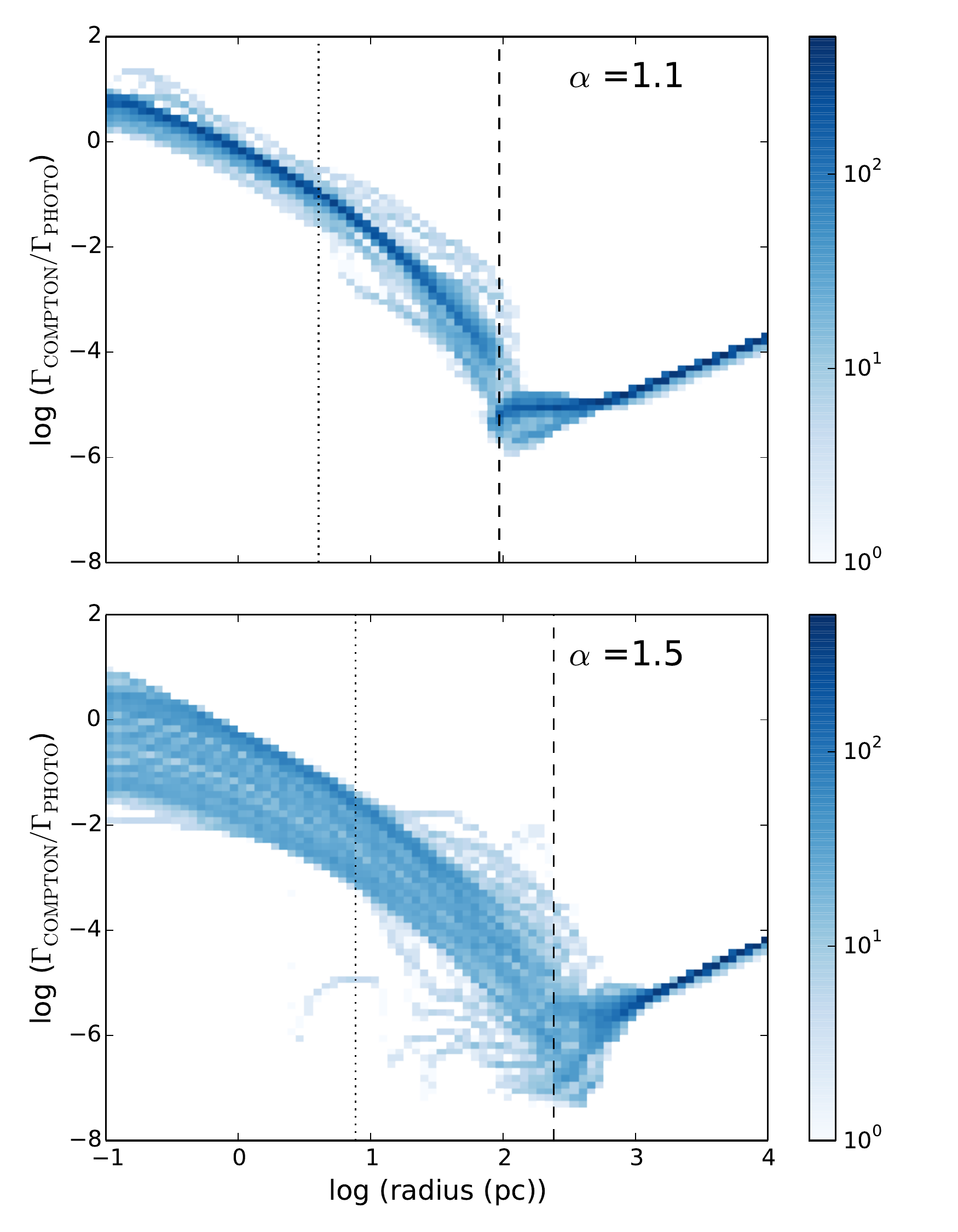}
\caption{2D histogram for the distribution of the ratio between 
Compton heating and photo-heating as a function of radius for the
spectral index $\alpha=1.1$ (top) and $1.5$ (bottom). The 
simulation with a hard spectrum (top panel) shows less variation
of the ratio compared to the one with a softer spectrum (bottom). Only at
small radius does Compton heating dominate photo-heating, whereas 
photo-heating is the primary source of heating for the most of the
volume of the \str~sphere. For $\alpha = 1.5$, Compton heating
affects the heating dominantly only at the peak luminosity due to
the strong oscillation of accretion rate. Vertical lines show the
I-front (dashed) and effective accretion radius (dotted) from
time-averaged profiles, respectively.
}
\label{fig:profile}
\end{figure}

Fig.~\ref{fig:profile} shows the distribution of the ratio between
the Compton and photo-heating rates as a function of radius for 
$\lmone$ model. Since the
ratio at each radius changes as a function of time (neutral fraction
of H and He changes depending on the accretion luminosity) due to
the oscillatory behaviour of accretion, we show a 2D histogram
of the combined outputs up to $135$~Myr with regular time-step. Note that the
relative distribution is relevant here since the number in each bin depends
on the bin size and the frequency of the simulation outputs. The top panel
shows the simulation with $\alpha=1.1$ and the bottom shows the
simulation with $\alpha = 1.5$. For both cases, Compton heating
becomes dominant as the radial position inside the \str~sphere
becomes nearer to the BH, while the ratio increases as a function
of radius outside the \str~sphere (shown as narrow distribution in
Fig.~\ref{fig:profile}). Note that both Compton and photo-heating
rates are low outside the \str~sphere due to the high optical depth.

\begin{figure}
\includegraphics[width=85mm]{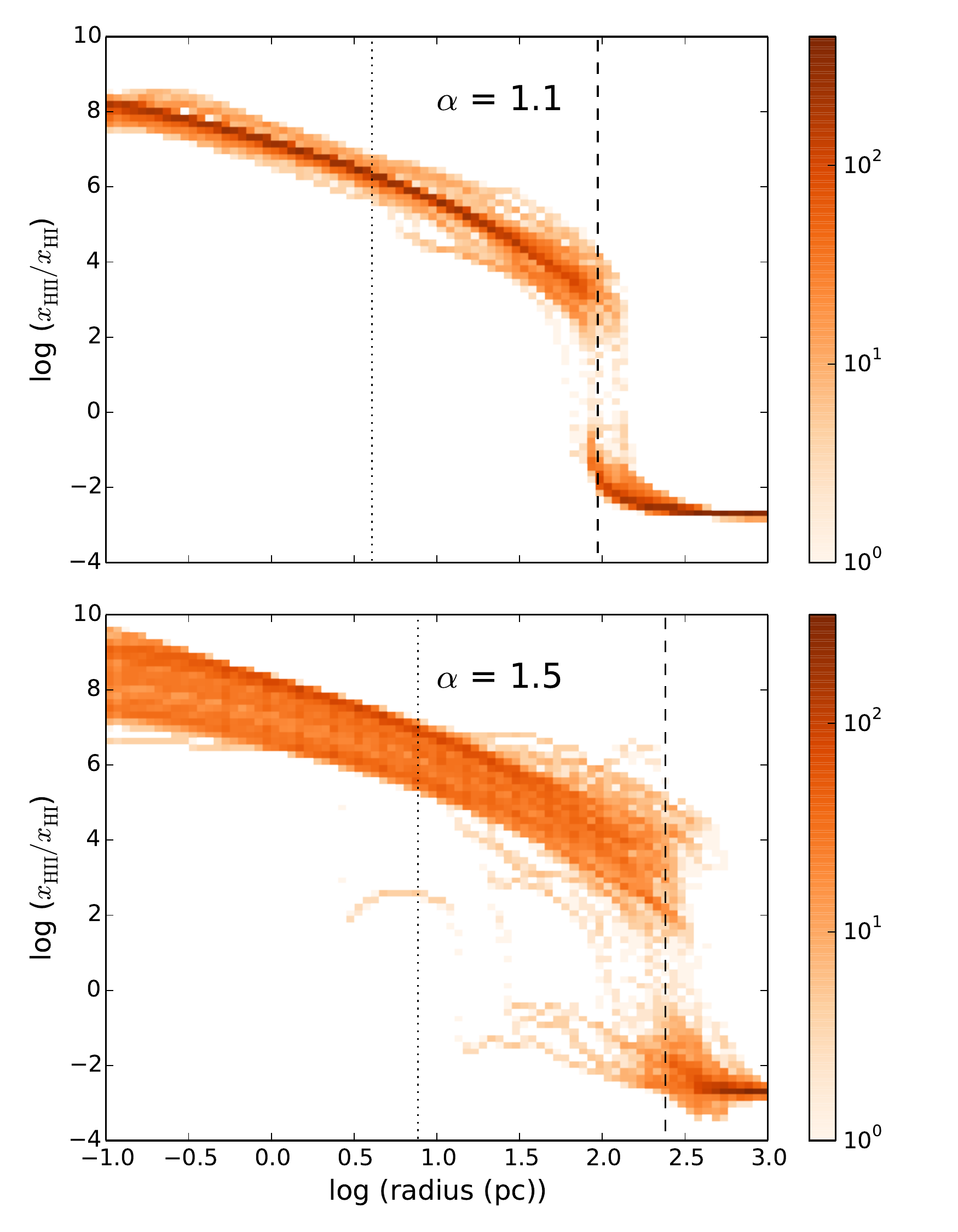}
\caption{2D histogram for the distribution of the ratio between the
abundances of \hii~and \hi~as a function of radius for the spectral
index $\alpha=1.1$ (top) and $1.5$ (bottom) for the same simulations shown
in Fig.~\ref{fig:profile}. The ratio between
Compton heating and photo-heating as a function of radius shown in
Fig.~\ref{fig:profile} can be well explained by the ratio between
the ionized and neutral fractions. Same vertical lines drawn in
Fig.~\ref{fig:profile} are shown.}
\label{fig:abundance}
\end{figure}

The ratio of the Compton-heating and photo-heating rates as a function
of radius can be explained directly by the ratio between neutral
and ionized fractions as mentioned in Sec~\ref{sec:intro}. 
Fig.~\ref{fig:abundance} shows the ratio distribution
between the abundances of \hii~and \hi~for the same simulations shown in
Fig.~\ref{fig:profile}. The Compton heating rate is proportional
to the electron number density as in Equation~(\ref{eq:compt_integral})
where the ionization fraction is close to unity inside
the \str~sphere. In contrast, the photo-heating rate
is proportional to the neutral fraction of the gas, which
increases as a function of radius inside the \str~sphere. This
fact explains well the radial behaviour of the relative strength of the
two heating mechanisms.

Only a tiny fraction of the \hii~region, more than 2 orders of
magnitude smaller in radius than the I-front, is dominated by the Compton
heating for $\alpha=1.1$. Compton heating becomes more efficient
at low $\alpha$ due to the suppressed oscillatory behaviour as
shown at the top panel of Fig.~\ref{fig:abundance}, whereas Compton
heating is dominant over photo-heating only during the peak
accretion phase for $\alpha = 1.5$ (bottom panel of Fig.~\ref{fig:abundance}).

\section{Summary and Discussion}
\label{sec:summary}
In this paper, we investigate the relative role of Compton-heating
and photo-heating on radiation-regulated accretion on
to BHs. We summarize our findings as follows.

\begin{itemize}
\item We find that the Compton heating is important when the spectrum
of the BH radiation is hard ($\alpha \sim 1$) or $T_C=3\times
10^7$~K. The oscillatory behaviour due to the radiative feedback
loop becomes weak and
the average accretion rate is as well suppressed to $0.01$ ($\lmone$
model)--$0.1$ ($\lmtwo$ model) per cent of the Bondi rate.
The accretion rate increases and the oscillatory behaviour becomes
stronger with increasing spectral index. When the spectrum is soft
($\alpha > 1.5$), photo-heating prevails regardless of models making
the effect of Compton heating negligible.

\item Compton heating is important near the BH inside the
\str~sphere, while the photo-heating still prevails in most part of
the \str~sphere. In particular, we emphasize that the photo-heating
dominates the Compton heating at the radial scale of the I-front where
the gas supply to the inside the \str~sphere is regulated.
The ratio of Compton-heating and photo-heating as a function
of radius is well explained by the ratio between the ionized and
neutral fraction.

\item The accretion feedback model $\lmtwo$ shows physically
consistent results with the radiatively inefficient ADAF model where
the accretion rate is low and the spectrum is hard ($\alpha \sim
1$). $\lmtwo$ model shows intermediate results between the feedback model
$\lmone$ and photo-heating-only model.  

\end{itemize}

We limit the results of the current study to the case of accretion
on to BHs from a neutral, dense, and warm medium with $T \sim 10^4~$K.
When the gas is hot, ionized and optically thin as found in elliptical
galaxies or at the central region of galaxy clusters, Compton
heating is regarded as the critical heating mechanism for the BH
feedback, especially when the spectrum is hard.

We fix the hardness of the spectrum during a simulation in the current work;
however, the state transition of the BH spectrum should play an
important role in determining the phenomena discussed here \citep{Gan:2014}. Most
of the gas accretion happens during the high luminosity phase where
the spectrum is {\it soft} while the spectrum stays in a ${\it hard}$
state during the low accretion phase \citep{Yuan:2009, Yuan:2014}. Dependence of the state
transition on the BH luminosity should be considered in future
work.

In this work, we assume an idealised power-law spectrum 
with a fixed energy range so that we can express the Compton
temperature in an analytic manner. However, the high-energy observation
of SMBHs reveals that the spectral energy distribution in the range
of UV and X-ray shows a deviation from the single power-law. 
\citep{Sazonov:2004} found a double-peaked distribution from the
weighted sum of obscured and unobscured quasar spectrum and calculated
the effective Compton temperature accurately $T_C= (1.9 \pm 0.8)
\times 10^7$~K. However they point out that the contribution of the
blue bump to the Compton heating is less than 25 per cent, which
is smaller than the error of the calculated Compton temperature.
Additionally, the approximate estimation of the effective spectral
slope of the averaged spectrum of the quasars in their work still
falls within the range of $1 < \alpha < 1.5$. Considering the
effective spectral slope and the Compton temperature, applying the
double-peaked spectrum to the simulations is not likely to produce
a significant qualitative difference from the current results.

\section*{Acknowledgements} The authors thank the referee, Luca
Ciotti for constructive comments and positive feedback. KP is
supported by the Urania E. Stott Fellowship of The Pittsburgh
Foundation. This research was supported
in part by the National Science Foundation under Grant No. NSF
PHY11-25915 via the program {\it A Universe of Black Holes} hosted
by the Kavli Institute for Theoretical Physics. MR's research is
supported by NASA grant NNX10AH10G, NSF CMMI1125285
and ILP LABEX (under reference ANR-10-LABX-63) supported by French
state funds managed by the ANR within the Investissements d'Avenir
programme under reference ANR-11-IDEX-0004-02. TDM acknowledges the
National Science Foundation, NSF Petapps, OCI-0749212 and NSF
AST-1009781 for support. CSR and MR thank the National Science Foundation
for support under the Theoretical and Computational Astrophysics
Network (TCAN) grant AST1333514. This paper benefits from the private
conversation with Jeremiah Ostriker. The numerical simulations in
this paper were performed using the cluster facilities (``ferrari"
and ``coma") of the McWilliams Center for Cosmology at Carnegie
Mellon University.

\bibliographystyle{mn2e}
\bibliography{park_bh}

\label{lastpage}

\end{document}